\documentstyle[12pt]{article}

\newcommand{\beq}{\begin{equation}}
\newcommand{\eeq}{\end{equation}}
\newcommand{\beqa}{\begin{eqnarray}}
\newcommand{\eeqa}{\end{eqnarray}}
\newcommand{\beqar}{\begin{eqnarray*}}
\newcommand{\eeqar}{\end{eqnarray*}}

\newcommand{\Ga}{\Gamma}

\newcommand{\inn}{\!\cdot\!}

\newcommand{\z}{\zeta}

\newcommand{\eg}{{\it e.g.,}\ }
\newcommand{\ie}{{\it i.e.,}\ }
\newcommand{\labell}[1]{\label{#1}} %{\label{#1}} %
\newcommand{\reef}[1]{(\ref{#1})}
\newcommand\prt{\partial}

\newcommand\tG{{\widetilde G}}

\newcommand\tV{{\widetilde V}}

\newcommand\Tr{{\rm Tr}}
\begin{document}

\thispagestyle{empty} \rightline{\small hep-th/0209068 \hfill
IPM/P-2002/038} \vspace*{1cm}

\begin{center}
{\bf \Large
On-shell S-matrix and tachyonic effective actions  \\

 }
\vspace*{1cm}

{Mohammad R. Garousi}\\
\vspace*{0.2cm}
{\it Department of Physics, Ferdowsi university, Mashhad, Iran}\\
\vspace*{0.1cm}
and\\
{\it Institute for Studies in Theoretical Physics and Mathematics
IPM} \\
{P.O. Box 19395-5531, Tehran, Iran}\\
\vspace*{0.4cm}

\vspace{2cm}
ABSTRACT
\end{center}
We evaluate various  disk level four-point functions involving
the massless scalar and tachyon
  vertex operators
in the presence of background B-flux in superstring theory. By
studying    these amplitudes in  specific limits, we find
couplings of two scalars with two tachyons, and couplings of four
tachyons on the world-volume of non-BPS D-branes of superstring
theory. They are fully consistent with the non-commutative
tachyonic Dirac-Born-Infeld  effective action. They also fix the
coefficient of  $T^4$ term in the expansion of the tachyon
potential  around its maximum.

\vfill
\setcounter{page}{0}
\setcounter{footnote}{0}
\newpage

\section{The idea} \label{intro}
Recently different tachyonic effective actions has been used to
describe the time evolution of unstable D-branes in string
theory\cite{sen1,gg,ssst,jm,nlis}\footnote{For early studies of
open string tachyon dynamics, see \cite{kb}.}. In particular, Sen
has shown that the string theory produces a pressure-less gas with
non-zero energy density at the late time of the tachyon
condensation\cite{sen1}. In this paper, he  showed that these
results can be derived also from the tachyonic
Dirac-Born-Infeld(DBI) effective action\cite{mg,ebmr} around its
true vacume. Other possible applications of this action to
cosmology have been discussed in \cite{cosmology}. This action
was claimed in \cite{mg}
 that  is  deducible   from on-shell string theory
scattering amplitudes. In this paper we would like to see to what
extent the string theory S-matrix elements produce this action.

In inferring effective action from string theory  on-shell
S-matrix elements,  one usually evaluates the  S-matrix elements
and then expands it in the limit $\alpha'\rightarrow 0$, \ie low
energy limit. Then one writes a low energy effective action in
the field theory that reproduces the leading  terms of the
expansion. For the case that the field theory involves only
massless fields, \ie flat potential,  this expansion is unique,
though many apparently different but physically identical low
energy actions may produce them. They may in turn be
 related to each other by some field
redefinition\cite{rmat}. However, if one includes  a tachyonic
field, then the corresponding limit of the S-matrix elements is
not $\alpha'\rightarrow 0$, in general, since the action around
the maximum of the tachyon potential itself is not low energy
effective action anymore. Different expansion of a S-matrix
element may then be correspond to physically different effective
actions.  For example, consider the S-matrix element of one
closed string tachyon and two open string massless scalars, \ie
flat potential,   in Type 0 theory\cite{ahik,mg2}\footnote{For
simplicity, we assumed that the closed string field does not
depend on the transverse coordinates.}, \beqa
A(\tau,X,X)&\sim&\z_1\inn
\z_2\frac{\Ga(-2s)}{\Ga(-s)\Ga(-s)}\,\,,\nonumber\eeqa where
$s=-2k_1\inn k_2$ and $\z^i,k^a$ are the scalar polarization and
momentum, respectively\footnote{Our convention sets $\alpha'=2$.
Our index conventions are that early Latin indices take values in
the world-volume, \eg $a,b=0,1,...,p$ , and middle Latin indices
take values in the transverse space, \eg $i,j=p+1,...,8,9$.}. The
low energy expansion, \ie $s\rightarrow 0$, is \beqa
A&\sim&\z_1\inn \z_2(k_1\inn k_2+O(s^2))\,\,.\nonumber\eeqa The
first term above is reproduced in field theory by the low energy
DBI action, and $O(s^2)$ terms are related to higher derivative
terms which are not included in the DBI action.

Now consider the S-matrix element of one closed string tachyon
and two open string tachyons  at the top of the open string
tachyon potential. A simple calculation yields
 \beqa
A(\tau,T,T)&\sim&\frac{\pi}{4}\frac{\Ga(-2s)}{\Ga(-s)\Ga(-s)}\,\,,
\label{att}\eeqa where $s=-1/2-2k_1\inn k_2$,  and $k^a$ is
momentum of the open string tachyon.  Expansion of this S-matrix
around $s=-1/2$
 is \beqa
A&\sim&\frac{1}{4}+2\ln(2)\,k_1\inn k_2+O((k_1\inn
k_2)^2)\,\,.\label{aexpand1}\eeqa The first two terms of this
expansion are reproduced exactly by the BSFT
action\cite{dkmm}\footnote{We have used the fact that the closed
string tachyon in Type 0 theory normalizes the D-brane tension as
$T_p\rightarrow T_p(1+\tau/4+\cdots)$\cite{mg3}, and we have kept
only linear term for the closed string tachyon.} \beqa
S&=&-T_p\int dx\, \frac{\tau}{4}
e^{-T^2/4}\frac{\sqrt{\pi}\Ga((\prt T)^2+1)}{\Ga((\prt T)^2+1/2)}
\,\,,\label{bac}\\
&=&-T_p\int dx\,\frac{\tau}{4}\left(1-\frac{T^2}{4}+2\ln(2)\,(\prt
 T)^2+\cdots\right)\,\,,\nonumber\eeqa and $O((k_1\inn k_2)^2)$
 terms are related to higher derivative terms,
\eg $\prt_a\prt_b T\prt^a\prt^b T$,
 which are not included in \reef{bac}. The expansion
 \reef{aexpand1} is also completely consistent with the sigma model
 effective action\cite{aat2}\footnote{Note that the convention in
 \cite{aat2} sets $\alpha'=1/(2\pi)$.}.
 Now, expansion of
 S-matrix element \reef{att}
around $s=0$ (the same limit as for  the scalar case)  yields
\beqa A&\sim&\frac{\pi}{8}(\frac{1}{2}+2k_1\inn k_2
+O(s^2))\,\,.\nonumber\eeqa The first two terms here are
reproduced exactly by the tachyonic DBI action\cite{mg,ebmr} \beqa
S&=&-T_p\int dx\,\frac{\tau}{4}
V(T)\sqrt{-\det(\eta_{ab}+4\pi\prt_a T\prt_b T)}\,\,,
\label{dbiac}\\
&=&-T_p\int dx\,\frac{\tau}{4}\left(1-\frac{\pi}{2}T^2+2\pi(\prt
 T)^2+\cdots\right)\,\,,\nonumber\eeqa
 where in the second line we have used the
 expansion $V(T)=1-\frac{\pi}{2}T^2+O(T^4)$ at the top of  the tachyon
 potential in this action. In this case, $O(s^2)$ terms may be
related to higher
derivative terms, \eg $T\prt_a\prt^a T$, which are not included
in \reef{dbiac}\footnote{The higher derivative terms may have
significant effect at the top of potential, however, according to
the results in \cite{sen1} they are not important at the minimum
of tachyon potential.}. Since actions \reef{bac} and \reef{dbiac},
around the top of their potentials, are consistent with the
leading terms of the S-matrix element \reef{att}  at different
limits, one may expect that they are not related to each other by
field redefinition. However, around  the minimum of their
potentials, they may be related to each other by some field
redefinition.

In this paper we would like to extend the above discussion to the
case of S-matrix elements of four tachyon or scalar vertex
operators in the superstring theory in the presence of background
B-flux . More specifically, we would like to find  appropriate
expansion for the S-matrix elements that  correspond  to the
tachyonic DBI  action\reef{dbiac}. As we will see in the text, the
background B-flux makes it easier to find the desired limit of the
amplitudes.  The analysis of S-matrix element of four tachyons
enables us to find the coefficient of $T^4$ in the tachyon
potential in this action.

In section 2.1, we recall the S-matrix element of four scalar
vertex operators. This amplitude has three Mandelstam variables
$s,t,u$ satisfying $s+t+u=0$, and has massless pole in $s$-, $t$-
and $u$-channels. In this case, the DBI action correspond to the
low energy expansion of the amplitude which is unique, \ie
$(s,t,u)\rightarrow 0$. In section 2.2, we evaluate the S-matrix
element of two scalar and two tachyon vertex operators. The
amplitude here has massless pole only in $t$-channel. The
Mandelstam variables in this case
 satisfy $s+t+u=-1/2$, so we expand the amplitude around
$t\rightarrow 0,\,(s,u)\rightarrow -1/4$. In section 2.3, we
evaluate the S-matrix element of four tachyon vertex operators.
The amplitude in this case has  massless pole  in all $s$-, $t$-,
 and
 $u$-channels. The
Mandelstam variables here satisfy $s+t+u=-1$. We expand
  the amplitude
 around $s\rightarrow 0,\,(t,u)\rightarrow -1/2$ to produce
 massless pole of field theory in $s$-channel, around
 $t\rightarrow 0,\, (s,u)\rightarrow -1/2$ to produce
 massless pole of field theory in $t$-channel, and around
 $u\rightarrow 0,\, (t,s)\rightarrow -1/2$ to produce the massless
 pole of theory in $u$-channel.
  In section 3, we compare the leading
 terms
of  the amplitudes  with the non-commutative version of the
tachyonic DBI action.

\section{Scattering calculation}\label{scatt}
In this section, using the world-sheet conformal field theory
technique, we evaluate various 4-point function of scalar and
tachyon vertex operators in the presence of background B-flux. All
these
amplitudes have massless pole, as well as an  infinite tower of
massive poles.
 Then we find
appropriate  limits which reduce the amplitudes to their massless
pole, and infinite number of contact terms.
 We begin with recalling  the scattering
amplitude of four massless scalar vertex operators.
\subsection{Four  scalars amplitude}
Scattering amplitude of four vector vertex operators in
superstring theory is evaluated in \cite{mgjs}, and its low energy
effective action is also studied, for example, in \cite{at}. To
find the amplitude corresponding to four scalar vertex operators,
one may use the result in \cite{mgjs} in which the vector
polarizations $\z^a$ are replaced by the scalar polarizations
$\z^i$. Since we are interested in the scattering amplitude in the
presence of B-flux, one should use $G=(1/(\eta+B))_S$ as the
world volume metric, and also should add an appropriate  phase
factor to the amplitudes in one cycling of the vertex operators
with the non-commutative parameter tensor
$\theta=(4\pi/(\eta+B))_A$ \cite{nsew}. Adding all non-cyclic
permutation of the vertex operators, one ends up with the
following amplitude: \beqa
A(\z_1,\z_2,\z_3,\z_4)&=&\,A_s(\z_1,\z_2,\z_3,\z_4)+\,A_u
(\z_1,\z_2,\z_3,\z_4)+
\,A_t(\z_1,\z_2,\z_3,\z_4)\,\,,
\nonumber\eeqa where $A_s,A_u$, and
$A_t$ are the part of the amplitude that  have  massless pole in
$s$-, $u$- and $t$-channels, respectively. They are
 \beqa A_s&= &-\frac{ic}{8\pi^2T_p}\z_1\inn\z_2\,\z_3\inn\z_4
\left(\frac{}{}\frac{\Ga(-2s)\Ga(1-2t)}{\Ga(-2s-2t)}(e^{i\pi
l_{12}+i\pi l_{34}}
+e^{i\pi l_{14}-i\pi l_{23}})\right.\nonumber\\
&&+\frac{\Ga(-2s)\Ga(1-2u)}{\Ga(-2s-2u)}(e^{i\pi l_{13}-i\pi
l_{24}}+e^{i\pi
l_{12}-i\pi l_{34}})\nonumber\\
&&\left.-\frac{\Ga(1-2t)\Ga(1-2u)}{\Ga(1-2t-2u)}(e^{i\pi
l_{14}+i\pi
l_{23}}+e^{i\pi l_{13}+i\pi l_{24}}) \frac{}{} \right)\,\,,
\nonumber\\
A_u&=&-\frac{ic}{8\pi^2T_p}\z_1\inn\z_3\,\z_2\inn\z_4\left(-\frac{}{}
\frac{\Ga(1-2s)\Ga(1-2t)}{\Ga(1-2s-2t)}(e^{i\pi l_{12}+i\pi
l_{34}}
+e^{i\pi l_{14}-i\pi l_{23}})\right.\nonumber\\
&&+\frac{\Ga(-2u)\Ga(1-2s)}{\Ga(-2u-2s)}(e^{i\pi l_{13}-i\pi
l_{24}}+e^{i\pi
l_{12}-i\pi l_{34}})\nonumber\\
&&\left.+\frac{\Ga(-2u)\Ga(1-2t)}{\Ga(-2u-2t)}(e^{i\pi l_{14}+i\pi
l_{23}}+e^{i\pi l_{13}+i\pi l_{24}}) \frac{}{}
\right)\,\,,\nonumber\\
A_t&=&-\frac{ic}{8\pi^2T_p}\z_1\inn\z_4\,\z_2\inn\z_3\left(\frac{}{}
\frac{\Ga(-2t)\Ga(1-2s)}{\Ga(-2t-2s)}(e^{i\pi l_{12}+i\pi l_{34}}
+e^{i\pi l_{14}-i\pi l_{23}})\right.\nonumber\\
&&-\frac{\Ga(1-2s)\Ga(1-2u)}{\Ga(1-2s-2u)}(e^{i\pi l_{13}-i\pi
l_{24}}+e^{i\pi
l_{12}-i\pi l_{34}})\nonumber\\
&&\left.+\frac{\Ga(-2t)\Ga(1-2u)}{\Ga(-2t-2u)}(e^{i\pi l_{14}+i\pi
l_{23}}+e^{i\pi l_{13}+i\pi l_{24}}) \frac{}{}
\right)\,\,,\nonumber\eeqa where
 $l_{ij}=k_i\inn \theta\inn k_j/(2\pi)$ and \beqa
s&=&-(k_1+k_2)^2\,=\,-k_1^2-k_2^2-2k_1\inn
k_2\,=\,-k_3^2-k_4^2-2k_3\inn k_4\,\,,\nonumber\\
t&=&-(k_2+k_3)^2\,=\,-k_2^2-k_3^2-2k_2\inn
k_3\,=\,-k_1^2-k_4^2-2k_1\inn k_4\,\,,\labell{mandel}
\\
u&=&-(k_1+k_3)^2\,=\,-k_1^2-k_3^2-2k_1\inn
k_3\,=\,-k_2^2-k_4^2-2k_2\inn k_4\,\,,\nonumber \eeqa where in the
present case $k_i^2=0$. We have also normalized the amplitude
here and in the subsequent sections by the factor $ic/(8\pi^2
T_p)$ where $c=\sqrt{-\det(\eta+B)}$. Note that each amplitude has
a term which has no massless pole. They produce only contact
terms upon expanding the amplitude. The amplitudes $A_s$, $A_u$
and $A_t$ are very similar, so we only analyze in some details the
$A_t$ amplitude. The low energy limit, \ie $(t,s,u)\rightarrow 0$
of the gamma functions in this amplitude are \beqa
\frac{\Ga(-2t)\Ga(1-2s)}{\Ga(-2t-2s)}&=&\frac{2u}{-2t}+
\frac{2\pi^2}{3}us+\cdots\,\,,\nonumber\\
\frac{\Ga(-2t)\Ga(1-2u)}{\Ga(-2t-2u)}&=&\frac{2s}{-2t}+
\frac{2\pi^2}{3}us+\cdots\,\,,\nonumber\\
\frac{\Ga(1-2s)\Ga(1-2u)}{\Ga(1-2s-2u)}&=&1-\frac{2\pi^2}{3}us+
\cdots\,\,,\nonumber\eeqa where we have used the on-shell
condition of $s+t+u=0$, and dots above represents terms with more
than four momenta. In field theory, using the non-commutative DBI
action, one finds the following massless
 $t$-channel amplitude (see \eg \cite{mg} for details): \beqa
A_t'&=&(\tV_{\phi_2\phi_3A})^a(\tG_A)_{ab}(\tV_{A\phi_1\phi_4})
^b\,\,,\nonumber\\
&=&\left(\frac{ic}{(2\pi)^2T_p}\right)\z_1\inn\z_4\,\z_2\inn\z_3
\,\frac{\sin(\pi
l_{23})\sin(\pi l_{14})(u-s)}{t}\,\,,\nonumber\\
&=&\frac{-ic}{8\pi^2T_p}\z_1\inn\z_4\,\z_2\inn\z_3\,\left((e^{i\pi
l_{12}+i\pi l_{34}}+e^{i\pi l_{14}-i\pi
l_{23}})(\frac{2u}{-2t}-\frac{1}{2})\right.\nonumber\\
&&\left.\qquad\qquad\qquad\qquad+(e^{i\pi l_{14}+i\pi
l_{23}}-e^{i\pi l_{13}+i\pi
l_{24}})(\frac{2s}{-2t}-\frac{1}{2})\right)\,\,.\nonumber\eeqa Now
subtracting the field theory massless pole from the string theory
scattering amplitude, one ends up with the following contact
terms in the $t$-channel: \beqa
A_t-A_t'&=&-\frac{ic}{8\pi^2T_p}\z_1\inn\z_4\,\z_2\inn\z_3
\left(\frac{}{}(e^{i\pi l_{12}+i\pi l_{34}} +e^{i\pi l_{14}-i\pi
l_{23}})(\frac{1}{2}+\frac{2\pi^2}{3}us+
\cdots)\right.\nonumber\\
&&+(e^{i\pi l_{13}-i\pi l_{24}}+e^{i\pi
l_{12}-i\pi l_{34}})(-1+\frac{2\pi^2}{3}us+\cdots)\nonumber\\
&&\left.+(e^{i\pi l_{14}+i\pi l_{23}}+e^{i\pi l_{13}+i\pi
l_{24}}) (\frac{1}{2}+\frac{2\pi^2}{3}us+\cdots)\frac{}{}
\right)\,\,.\label{att'}\eeqa Doing the same analysis for the
$A_s$ and $A_u$, one finds the following total contact terms:\beqa
A_c(\z_1,\z_2,\z_3,\z_4)&=&A_c^0(\z_1,\z_2,\z_3,\z_4)+
A_c^4(\z_1,\z_2,\z_3,\z_4)+
\sum_{n>4}A_c^n(\z_1,\z_2,\z_3,\z_4)\,\,,\nonumber\eeqa where
$A_c^0$ contains, apart from the phase factor, contact terms with
no momentum and is zero when the background B-flux vanishes, \beqa
A_c^0&\!\!=\!\!&\frac{-ic}{8\pi^2T_p}\left(e^{i\pi l_{12}+i\pi
l_{34}} +e^{i\pi l_{14}-i\pi
l_{23}}\right)\left(\frac{1}{2}\z_1\inn\z_4\,\z_2\inn\z_3
+\frac{1}{2}\z_1\inn\z_2\,\z_3\inn\z_4-\z_1\inn\z_3\,
\z_2\inn\z_4\right)\nonumber\\
&&-\frac{ic}{8\pi^2T_p}\left(e^{i\pi l_{13}-i\pi l_{24}} +e^{i\pi
l_{12}-i\pi l_{34}}\right)\left(-\z_1\inn\z_4\,\z_2\inn\z_3
+\frac{1}{2}\z_1\inn\z_2\,\z_3\inn\z_4+
\frac{1}{2}\z_1\inn\z_3\,\z_2\inn\z_4\right)\nonumber\\
&&-\frac{ic}{8\pi^2T_p}\left(e^{i\pi l_{14}+i\pi l_{23}} +e^{i\pi
l_{13}+i\pi
l_{24}}\right)\left(\frac{1}{2}\z_1\inn\z_4\,\z_2\inn\z_3
-\z_1\inn\z_2\,\z_3\inn\z_4
+\frac{1}{2}\z_1\inn\z_3\,\z_2\inn\z_4\right). \labell{ac01}\eeqa
$A_c^4$ contains terms with four momenta which are non vanishing
even in the $B=0$ limit. They are
 \beqa A_c^4&\!\!=\!\!&
\frac{-ic}{6T_p}\left(e^{i\pi l_{12}+i\pi l_{34}} +e^{i\pi
l_{14}-i\pi l_{23}}+e^{i\pi l_{13}-i\pi l_{24}} +e^{i\pi
l_{12}-i\pi l_{34}}+e^{i\pi l_{14}+i\pi l_{23}} +e^{i\pi
l_{13}+i\pi
l_{24}}\right)\nonumber\\
&&\times\left(\frac{}{}\z_1\inn\z_4\,\z_2\inn\z_3\left(\frac{}{}(k_2
\inn
k_3)(k_1\inn k_4) -(k_1\inn k_2)(k_3\inn k_4)-(k_1\inn
k_3)(k_2\inn k_4)\frac{}{}\right)\right.
\nonumber\\
&&+\z_1\inn\z_2\,\z_3\inn\z_4\left(\frac{}{}(k_1\inn k_2)(k_3\inn
k_4)-(k_2\inn
k_3)(k_1\inn k_4)-(k_1\inn k_3)(k_2\inn k_4)\frac{}{}\right)
\nonumber\\
&&\left. +\z_1\inn\z_3\,\z_2\inn\z_4\left(\frac{}{}(k_1\inn
k_3)(k_2\inn k_4) -(k_1\inn k_2)(k_3\inn k_4)-(k_2\inn
k_3)(k_1\inn
k_2)\frac{}{}\right)\frac{}{}\right)\,.\labell{ac41}\eeqa And
$A_c^n$ with $n>4$ contains contact terms with more than four
momenta. They are related to higher derivative terms which are
not included in the low energy DBI action. In writing above result
we have used the condition $s+t+u=0$ and \reef{mandel}.  In this
form as opposed to the form in \reef{att'}, each momentum appears
only once. This form is comparable with the DBI action which has
first derivative of scalar fields. We shall compare above contact
terms with the non-commutative DBI action in section 3.

\subsection{Two scalars and two tachyons amplitude}
The amplitude describing scattering of two tachyons to two
scalars is given by the following correlation: \beqa
A^{1234}&\sim&\int
dx_1dx_2dx_3dx_4\nonumber\\
&&\times\langle:V^T_{0}(2V\inn k_1,x_1): V^X_{-1}(\z_2,2V\inn
k_2,x_2):V^X_{-1}(\z_3,2V\inn k_3,x_3):V^T_0(2V\inn
k_4,x_4):\rangle\,,\nonumber\eeqa where we have used the doubling
trick to deal with only world sheet analytic fields\cite{mg4}. The
matrix $V$ is related to world volume metric and the
non-commutative parameter as $V_S=G$ and $ V_A=\theta/(4\pi)$
where the subscripts $S$ and $A$ mean symmetric and antisymmetric
 part of $V$ matrix, respectively.
 In above equation $k_1^2=k_4^2=1/4$, $k_2^2=k_3^2=0$, and
\beqa V^X_{-1}(\z,k,x)&=&e^{-\phi}\z_i\psi^ie^{ik\cdot X}\,\,,
\nonumber\\
V^T_0(k,x)&=&ik\inn\psi e^{ik\cdot X}\,\,.\nonumber \eeqa Using
the appropriate world sheet propagators, one can easily  evaluates
all the correlators above and show that the integrand is SL(2,R)
invariant. Removing  this symmetry from the integral by fixing the
position of three vertex operators at $x_1=0,x_3=1$, and $
x_4=\infty$, one finds \beqa A^{1234}&\sim&-4k_1\inn
k_4\z_2\inn\z_3\int_0^1dx\,x^{4k_1\cdot k_4-1}(1-x)^{4k_4\cdot
k_3}e^{i\pi l_{12}+i\pi l_{34}} \,\,.\nonumber\eeqa Adding all
non-cyclic permutation  of  the vertex operators, one finds that
the amplitude has only massless pole in $t$-channel, that is,
 \beqa A(\z_2,\z_3)\,=\,A_t(\z_2,\z_3)&= &
-\frac{ic}{8\pi^2T_p}\z_2\inn\z_3\left(\frac{}{}\frac{\Ga(-2t)\Ga(1/2-
2s)}
 {\Ga(-1/2-2t-2s)}(e^{i\pi
l_{12}+i\pi l_{34}}
+e^{i\pi l_{14}-i\pi l_{23}})\right.\nonumber\\
&&-\frac{\Ga(1/2-2s)\Ga(1/2-2u)}{\Ga(-2s-2u)}(e^{i\pi l_{13}-i\pi
l_{24}}+e^{i\pi
l_{12}-i\pi l_{34}})\nonumber\\
&&\left.+\frac{\Ga(-2t)\Ga(1/2-2u)}{\Ga(-1/2-2t-2u)}(e^{i\pi
l_{14}+i\pi l_{23}}+e^{i\pi l_{13}+i\pi l_{24}}) \frac{}{}
\right)\,\,,\label{atone}\eeqa where $s,t,u$ defined in
\reef{mandel}, and they satisfy the on-shell condition
$s+t+u=-1/2$. Now question is what is the limit that one   has to
take to produce the tachyonic DBI  effective action? The field
theory produces the following massless pole in the $t$-channel:
\beqa
A_t'&=&(\tV_{\phi_2\phi_3A})^a(\tG_A)_{ab}(\tV_{AT_1T_4})
^b\,\,,\nonumber\\
&=&\left(\frac{ic}{(2\pi)^2T_p}\right)\z_2\inn\z_3\,\frac{\sin(\pi
l_{23})\sin(\pi l_{14})(u-s)}{t}\,\,,\nonumber\\
&=&\frac{-ic}{8\pi^2T_p}\z_2\inn\z_3\,\left((e^{i\pi l_{12}+i\pi
l_{34}}+e^{i\pi l_{14}-i\pi
l_{23}})(\frac{1/2+2u}{-2t}-\frac{1}{2})\right.\nonumber\\
&&\left.\qquad\qquad\qquad\qquad+(e^{i\pi l_{14}+i\pi
l_{23}}-e^{i\pi l_{13}+i\pi
l_{24}})(\frac{1/2+2s}{-2t}-\frac{1}{2})\right)\,.\label{at'one}\eeqa
This pole is reproduced by the string theory amplitude
\reef{atone} if one takes for $t$ the limit
 $t\rightarrow 0$, and for $s,u$
 the symmetric limit $(s,u)\rightarrow -1/4$. Expansion of the gamma
functions
in \reef{atone} in this limit are
\beqa
\frac{\Ga(-2t)\Ga(1/2-2s)}{\Ga(-1/2-2t-2s)}&=&\frac{1/2+2u}{-2t}+
\frac{2\pi^2}{3}(u+1/4)(s+1/4)+\cdots\,\,,\nonumber\\
\frac{\Ga(-2t)\Ga(1/2-2u)}{\Ga(-1/2-2t-2u)}&=&
\frac{1/2+2s}{-2t}+\frac{2\pi^2}{3}(u+1/4)(s+1/4)
+\cdots\,\,,\nonumber\\
\frac{\Ga(1/2-2s)\Ga(1/2-2u)}{\Ga(-2s-2u)}&=&1-\frac{2\pi^2}{3}
(u+1/4)(s+1/4)+\cdots\,\,.\nonumber\eeqa Inserting above
expansion  into \reef{atone}, one finds that the string theory
amplitude produces exactly the field theory massless pole
\reef{at'one} in the above mentioned limit.

 To find the contact terms, one should
subtract the field theory massless pole \reef{at'one} from the
string theory amplitude \reef{atone}. Like in previous section,
they can be written as \beqa
A_c(\z_2,\z_3)&=&A_c^0(\z_2,\z_3)+A_c^4(\z_2,\z_3)+\sum_{n>4}A_c^n
(\z_2,\z_3)\,\,,
\label{ac}\eeqa where $A_c^0$  includes the contact terms that
have no momentum and is zero when the background field vanishes,
\beqa
A_c^0&=&-\frac{ic}{8\pi^2T_p}\z_2\inn\z_3\left(\frac{1}{2}\left(e^
{i\pi
l_{12}+i\pi l_{34}} +e^{i\pi l_{14}-i\pi
l_{23}}\right)\right.\nonumber\\
&&\left.-\left(e^{i\pi l_{13}-i\pi l_{24}} +e^{i\pi l_{12}-i\pi
l_{34}}\right)+\frac{1}{2}\left(e^{i\pi l_{14}+i\pi l_{23}}
+e^{i\pi l_{13}+i\pi l_{24}}\right)\right)\,\,.\labell{ac02} \eeqa
Using the fact that
  $u+1/4=-2k_1\inn k_3$ and $s+1/4=-2k_1\inn k_2$, one finds that
$A_c^4$
contact terms are proportional to $(k_1\inn k_2)(k_1\inn k_3)$.
 However,  the contact terms
 in this form is not appropriate
to compare with an action which has only first derivative terms,
since $k_1$
appears with power two and $k_4$ does not appear at all. They can be
rewritten in the suitable form using
 the on-shell condition $s+t+u=-1/2$, and
\reef{mandel},
 \beqa
A_c^4&\!\!=\!\!& \frac{-ic}{6T_p}\left(e^{i\pi l_{12}+i\pi
l_{34}} +e^{i\pi l_{14}-i\pi l_{23}}+e^{i\pi l_{13}-i\pi l_{24}}
+e^{i\pi l_{12}-i\pi l_{34}}+e^{i\pi l_{14}+i\pi l_{23}} +e^{i\pi
l_{13}+i\pi
l_{24}}\right)\nonumber\\
&&\times\z_2\inn\z_3\left((k_2\inn k_3)(k_1\inn k_4)-(k_1\inn
k_2)(k_3\inn k_4)-(k_1\inn k_3)(k_2\inn k_4)+\frac{1}{4}(k_2\inn
k_3) \right)\,.\labell{ac42}\eeqa In this   form each momentum
appears at most once which is in appropriate form for comparing
with the tachyonic DBI  action which includes at most the
 first derivative of
scalars and tachyon. The contact terms
 $A_c^n$ with $n>4$ in \reef{ac} contains all other
 terms.  One may expect that
 they are  related to higher derivative terms
that are not included in the tachyonic DBI action. We find a
justification for this in section 3.

\subsection{Four tachyons amplitude}
The amplitude describing scattering of two tachyons to two
tachyons is given by the following correlation: \beqa
A^{1234}&\sim&\int
dx_1dx_2dx_3dx_4\nonumber\\
&&\times\langle:V^T_{-1}(2V\inn k_1,x_1): V^T_{-1}(2V\inn
k_2,x_2):V^T_0(2V\inn k_3,x_3):V^T_0(2V\inn
k_4,x_4):\rangle\,,\nonumber\eeqa where $k_i^2=1/4$, and
\beqa V^T_{-1}(k,x)&=&e^{-\phi}e^{ik\cdot X}\,\,,\nonumber\\
V^T_0(k,x)&=&ik\inn\psi e^{ik\cdot X}\,\,.\nonumber \eeqa Here
again it is easy to evaluate all the correlators in the scattering
amplitude above and to show the integrand is SL(2,R) invariant.
Removing this symmetry by fixing $x_1=0,x_3=1$, and $x_4=\infty$,
one finds \beqa A^{1234}&\sim&-4k_3\inn
k_4\int_0^1dx\,x^{4k_1\cdot k_2-1}(1-x)^{4k_2\cdot k_3}e^{i\pi
l_{12}+i\pi l_{34}} \,\,.\nonumber\eeqa Adding all non-cyclic
permutations of the vertex operators, one finds that the whole
amplitude has massless pole in all $s$-, $t$- and $u$-channels,
that is, \beqa A&=&\,A_s+\,A_u+ \,A_t\,\,.\nonumber\eeqa However,
in this case the amplitudes in all channels are identical \ie
$A_s=A_t=A_u$, and
 \beqa
A_t&=&-\frac{ic}{8\pi^2T_p}\left(\frac{}{}
\frac{\Ga(-2t)\Ga(-2s)}{\Ga(-1-2s-2t)}(e^{i\pi
l_{12}+i\pi l_{34}}
+e^{i\pi l_{14}-i\pi l_{23}})\right.\nonumber\\
&&-\frac{\Ga(-2s)\Ga(-2u)}{\Ga(-1-2s-2u)}(e^{i\pi l_{13}-i\pi
l_{24}}+e^{i\pi
l_{12}-i\pi l_{34}})\nonumber\\
&&\left.+\frac{\Ga(-2t)\Ga(-2u)}{\Ga(-1-2t-2u)}(e^{i\pi
l_{14}+i\pi l_{23}}+e^{i\pi l_{13}+i\pi l_{24}}) \frac{}{}
\right)\,\,,\labell{atttt}\eeqa where
 $s,t,u$ satisfy the on-shell condition $ s+t+u=-1$.
 Here again to produce the $t$-channel
massless pole in field theory one has to take the limit
 $t\rightarrow 0$ for $t$, and  the symmetric limit
$(s,u)\rightarrow -1/2$ for $s,u$. The gamma functions in $A_t$ in
this limit
are
 \beqa
\frac{\Ga(-2t)\Ga(-2s)}{\Ga(-1-2t-2s)}&=&\frac{1+2u}{-2t}+
\frac{2\pi^2}{3}(u+1/2)(s+1/2)+\cdots\,\,,\nonumber\\
\frac{\Ga(-2t)\Ga(-2u)}{\Ga(-1-2t-2u)}&=&\frac{1+2s}{-2t}+
\frac{2\pi^2}{3}(u+1/2)(s+1/2)+\cdots\,\,,\nonumber\\
\frac{\Ga(-2s)\Ga(-2u)}{\Ga(-1-2s-2u)}&=&1-\frac{2\pi^2}{3}(u+1/2)
(s+1/2)+
\cdots\,\,.\nonumber\eeqa It is important to note that even though
$\Ga(-2s)$ and $\Ga(-2u)$ in above equations have massless pole
in $s$- and $u$-channels, respectively, they produce only contact
terms in the limit $(s,u)\rightarrow -1/2$ taken for $A_t$.
 The field
theory result for massless $t$-channel pole is \beqa
A_t'&=&(\tV_{T_2T_3A})^a(\tG_A)_{ab}(\tV_{AT_1T_4})^b\,\,,\nonumber\\
&=&\left(\frac{ic}{(2\pi)^2T_p}\right)\frac{\sin(\pi
l_{23})\sin(\pi l_{14})(u-s)}{t}\,\,,\nonumber\\
&=&\frac{-ic}{8\pi^2T_p}\left((e^{i\pi l_{12}+i\pi
l_{34}}+e^{i\pi l_{14}-i\pi
l_{23}})(\frac{1+2u}{-2t}-\frac{1}{2})\right.\nonumber\\
&&\left.\qquad\qquad\qquad\qquad+(e^{i\pi l_{14}+i\pi
l_{23}}-e^{i\pi l_{13}+i\pi
l_{24}})(\frac{1+2s}{-2t}-\frac{1}{2})\right)\,\,.\nonumber\eeqa
Subtracting the above field theory amplitude from the string
theory amplitude\reef{atttt}, one finds
 \beqa
A_t-A_t'&=&-\frac{ic}{8\pi^2T_p}\left(\frac{}{}(e^{i\pi
l_{12}+i\pi l_{34}}
+e^{i\pi l_{14}-i\pi l_{23}})(\frac{1}{2}+
\frac{2\pi^2}{3}(u+1/2)(s+1/2)+\cdots)\right.\nonumber\\
&&+(e^{i\pi l_{13}-i\pi l_{24}}+e^{i\pi
l_{12}-i\pi l_{34}})(-1+\frac{2\pi^2}{3}(u+1/2)(s+1/2)+\cdots)
\nonumber\\
&&\left.+(e^{i\pi l_{14}+i\pi l_{23}}+e^{i\pi l_{13}+i\pi l_{24}})
(\frac{1}{2}+\frac{2\pi^2}{3}(u+1/2)(s+1/2)+\cdots)\frac{}{}
\right)\,\,.\nonumber\eeqa For $A_s$ amplitude, one has to take
the limit $s\rightarrow 0$,  $(t,u)\rightarrow -1/2$, and
similarly for $A_u$ one has to take the limit $u\rightarrow 0$,
$(s,t)\rightarrow -1/2$ to produce the field theory massless pole
in $u$-channel . Adding the contact terms in all channels, one
finds the following total contact terms: \beqa
A_c&=&A_c^0+A_c^4+\sum_{n>4}A_c^n\,\,,\nonumber\eeqa where in this
case $A_c^0\,=\,0 $,  as  expected(see section 3),
 $A_c^4$ which has four momenta can be written in the following
form:
 \beqa A_c^4&=& -\frac{ic}{6T_p}\left(e^{i\pi
l_{12}+i\pi l_{34}} +e^{i\pi l_{14}-i\pi l_{23}}+e^{i\pi
l_{13}-i\pi l_{24}} +e^{i\pi l_{12}-i\pi l_{34}}+e^{i\pi
l_{14}+i\pi l_{23}} +e^{i\pi l_{13}+i\pi
l_{24}}\right)\nonumber\\
&&\qquad\qquad\qquad\times\left(-(k_2\inn k_3)(k_1\inn k_4)
-(k_1\inn k_2)(k_3\inn k_4)-(k_1\inn k_3)(k_2\inn k_4)+
\frac{1}{16}\right)\,,\labell{ac43}\eeqa where  we have used the
on-shell condition $s+t+u=-1$ and \reef{mandel} to write the
momenta in this amplitude in the form that   each momentum
appears at most once. And
 $A_c^n$ with $n>4$ contains all other
infinite contact terms. They may be  related to higher derivative
terms that are not included in the tachyonic DBI action. In next
section we shall compare $A_c^0$'s and $A_c^4$'s with the
tachyonic DBI action.

\section{Effective action}
In \cite{mg}, the S-matrix element of two tachyons and one
massless closed string $NSNS$ state in the presence of background
B-flux was  examined in details in order to find a tachyonic
effective action.
 The string theory amplitude in this simple case
has only one Mandelstam variable, $s$, and it also has massless
pole as well as infinite number of massive poles. The  only limit
that
 reduces the amplitude to its massless pole and
infinite number of contact terms is the limit $s\rightarrow 0$.
In this limit, the contact terms of the S-matrix element contain
$*'$ which indicates that the effective action, like the low
energy non-commutative DBI action,  should also have Wilson line
\cite{mg2,hl}.  Then it was shown that the massless pole and the
leading  contact term of the amplitude  are fully consistent with
the non-commutative version of the following tachyonic DBI
action\cite{mg, ebmr}\footnote{We explicitly restore $\alpha'$ in
this section.} :
 \beqa
S&=&-T_p\int d^{p+1}\sigma V(T)
e^{-\Phi}\sqrt{-\det(P[g_{ab}+b_{ab}]+2\pi\alpha'F_{ab}+2
\pi\alpha'\prt_a
T\prt_b T)} \,\,,\labell{dbiac2}\eeqa where
$V(T)=1-\frac{\pi}{2}T^2+O(T^4)$, $g_{ab}$ is flat space metric
$\eta_{ab}$ plus  its graviton fluctuation. Here $b_{ab},\Phi,A_a$
and $T$ are the antisymmetric Kalb-Ramond tensor, dilaton, gauge
field and the tachyon fluctuations, respectively. In above action
$P[\cdots]$ is also the pull-back of the closed string fields. For
 example,  $P[\eta_{ab}]=\eta_{ab}+\prt_a X^i\prt_b X_i$ in the
static gauge.

The non-abelian extension of this action is\cite{mg} \beqa
S&=&-T_p\int
d^{p+1}\sigma \Tr\left(V(T)\sqrt{\det(Q^i{}_j)}\right.\nonumber\\
&&\times\left.
e^{-\Phi}\sqrt{-\det(P[E_{ab}+E_{ai}(Q^{-1}-\delta)^{ij}E_{jb}]
+2\pi\alpha'F_{ab}+T_{ab})} \right)\,\,,\labell{nonab} \eeqa where
$E_{\mu\nu}=g_{\mu\nu}+b_{\mu\nu}$. The  indices in this action
are raised  and lowered by $E^{ij}$ and $E_{ij}$, respectively.
The matrices $Q^i{}_j$ and $T_{ab}$ are  \beqa
Q^i{}_j&=&\delta^i{}_j-\frac{i}{2\pi\alpha'}[X^i,X^k]E_{kj}
-\frac{1}{2\pi\alpha'}[X^i,T][X^k,T]E_{kj}\,\,,\nonumber\\
T_{ab}&=&2\pi\alpha'D_aTD_bT+D_aT[X^i,T](Q^{-1})_{ij}[X^j,T]
D_bT\nonumber\\
&&+iE_{ai}(Q^{-1})^i{}_j[X^j,T]D_bT+iD_aT[X^i,T](Q^{-1})_i{}^jE_{jb}
\nonumber\\
&&+iD_aX^i(Q^{-1})_{ij}[X^j,T]D_bT-iD_aT[X^i,T](Q^{-1})_{ij}
D_bX^j\,\,.
\nonumber\eeqa The trace in the action \reef{nonab} should be
completely symmetric between all non-abelian expression of the
form $F_{ab},D_aX^i,[X^i,X^j],D_aT,[X^i,T]$, individual $T$ of the
tachyon potential and individual $X^i$ of the Taylor expansion of
the closed string fields in the action\cite{mg5}. All world volume
derivatives in this action are covariant derivatives.

The abelian non-commutative  action when there is a background
B-flux with non-vanishing component only in the world volume
directions \ie $B_{ab}\neq 0$, and when the closed string fields
are constant \ie no closed string quantum fluctuation,  is the
same as the non-abelian action \reef{nonab} in which the world
volume  metric should be replaced by $G=(1/(\eta+B)_S$,  the
ordinary multiplication should be replaced by $*$-product with
non-commutative parameter $\theta=(2\pi\alpha'/(\eta+B))_A$, and
$T_p\rightarrow cT_p$\cite{nsew}. Note that the symmetrized trace
in non-abelian case transforms to symmetrized $*$-product in the
abelian non-commutative action.

Now it is straightforward to expand \reef{nonab} to find
different couplings in this action. The couplings between four
non-commutative scalar fields is \beqa {\cal{L}}(X,X,X,X)&=&-cT_p
\left(\frac{1}{16\pi^2\alpha'^2}[X^i,X^k][X_k,X_i]\right.\labell
{lpppp}\\
&&\left.-\frac{1}{4}(\prt_aX^i\prt_bX_i)(\prt^bX_j\prt^aX^j)+
\frac{1}{8}(\prt_aX^i\prt^aX_i)^2\right)\,\,, \nonumber\eeqa where
the multiplication is the symmetrized $*$-product. Using the
rescaling $X^i\rightarrow \frac{1}{\sqrt{T_p}}X^i$, it is easy to
verify that the coupling in the first line produce the contact
terms \reef{ac01} in momentum space. The symmetrized $*$-product
for the coupling in the first line  above is trivial, whereas it
has nontrivial effect on the terms in the second line.  Using
this,
 one can show that they produce exactly the contact terms in
\reef{ac41}. All these confirm the consistency between string
theory scattering amplitude at low energy and the low energy
non-commutative DBI action with symmetrized $*$-product.
 Now lets examine the coupling
between two scalars and two tachyons. Using the symmetrized
product, one finds only the following couplings: \beqa
{\cal{L}}(X,X,T,T)&=&-cT_p
\left(\frac{1}{4\pi\alpha'}[X^i,T][X_i,T]-\frac{\pi}{4}T^2
(\prt_aX^i\prt^aX_i)
\right.\labell{lpptt}\\
&&\left.-\pi\alpha'(\prt_aX^i\prt_bX_i)(\prt^bT\prt^aT)+
\frac{\pi\alpha'}{2}(\prt_aX^i\prt_aX_i)(\prt_bT\prt^bT)\right)\,\,,
\nonumber\eeqa where again the multiplication rule  is the
symmetrized $*$-product, and the world volume indices are raised
by $G^{ab}$. Now using the rescaling $T\rightarrow
\frac{1}{\sqrt{2\pi\alpha' T_p}}T$, one can show that the first
term in the first line above, reproduce the contact terms in
\reef{ac02},  the second term in the first line above produces
the last contact term in \reef{ac42}, and the other couplings in
the second line above produce the contact terms in \reef{ac42}
which have  four momenta. This confirms that the coupling
extracting from the tachyonic DBI
 action are consistent with the string theory scattering
amplitude at the limit $t\rightarrow 0,\,(s,u)\rightarrow -1/4$.
This also justifies the  ignoring  of  the $A_c^n$ terms with
$n>4$, \ie they are related to higher derivative terms that are
not included in the tachyonic DBI action \reef{nonab}.
 Finally, the coupling of four tachyons
extracted from \reef{nonab} is \beqa
{\cal{L}}(T,T,T,T)&=&-cT_p\left( \beta
T^4-\frac{\pi^2\alpha'}{2}T^2(\prt_aT\prt^aT)-
\frac{\pi^2\alpha'^2}{2}(\prt_aT\prt^aT)^2\right)\,\,,
\labell{ltttt}\eeqa where again the product is the symmetrized
$*$-product and the constant $\beta$ is the yet unknown
coefficient of $T^4$ in the tachyon potential. Here once again
one can observe that the terms with four derivatives reproduce
exactly the four momentum contact terms in \reef{ac43}, and the
rest produce the contact terms with no momentum in \reef{ac43}
provided $\beta=\pi^2/8$.

 The tachyon potential expanded around its
maximum, \ie around $T_{max}=0$, is then \beqa
V(T)=1-\frac{\pi}{2}T^2+\frac{\pi^2}{8}T^4+O(T^6)\,\,.\labell{finalv}
\eeqa The coefficient of $T^4$ is consistent with  Sen's
conjecture that the tachyon potential should have a minimum as
well\cite{sen2}. In fact if we ignore the $O(T^6)$ terms in the
potential, $V(T)$ has minimum at $T_{min}=\sqrt{2/\pi}$, and the
minimum of the potential is $V(T_{min})=0.5$ which is different
from the actual value of zero. This means that the higher order
terms $O(T^6)$ should be included in the potential. In fact  Sen
has shown that  the actual tachyon potential in the action
\reef{dbiac2} has minimum at $T_{min}\rightarrow \infty$, and the
behavior of the potential around the minimum should be like
$e^{-\sqrt{\pi}T}$\cite{sen3}\footnote{Note that the convention
in \cite{sen3} sets $\alpha'=1$.}. One may try to find a tachyon
potential with this
 behavior at  $T_{min}\rightarrow \infty$,
and expansion \reef{finalv} at $T_{max}=0$. For example the
following function \beqa
V(T)&=&\frac{1}{2\cosh(\sqrt{\pi}T)}\left(1+
\frac{1}{1+\pi^2T^4/6}\right)\,\,,\nonumber\eeqa has the correct
behaviors at minimum and maximum of the potential. It would be
interesting to extend the method used in this paper to find the
$O(T^6)$ terms,  and consequently to find the actual form of the
tachyon potential in the tachyonic DBI action\reef{dbiac2}. It
would be also interesting to apply this method to
 the bosonic string theory S-matrix elements which have both tachyonic
 and
  massless poles.

  {\bf Acknowledgement}: I would like to thank A. A. Tseytlin for
comments.

%\newpage

\end{document}